\makeatletter
\def\input@path{{elsevier-style/}}
\makeatother
\documentclass[a4paper,fleqn]{cas-dc}

\usepackage[numbers]{natbib}

\usepackage{tcolorbox}

\usepackage{tikz-cd}      

\begin{document}
\let\WriteBookmarks\relax
\def\floatpagepagefraction{1}
\def\textpagefraction{.001}
\shorttitle{The causal uncertainty principle}
\shortauthors{D. D. Reidpath}

\title [mode = title]{The causal uncertainty principle}

\author[1]{Daniel D. Reidpath}[type=author,
                        orcid=0000-0002-8796-0420]
\cormark[1]
\ead{dreidpath@qmu.ac.uk}

\affiliation[1]{organization={Institute for Global Health and Development, Queen Margaret University},
                city={Edinburgh},
               postcode={EH21 6UU}, 
                state={Scotland},
                country={United Kingdom}}

\cortext[cor1]{Corresponding author}

\begin{abstract}
\noindent\textbf{Background:} Internal and external validity stand in tension. Studies with strong causal identification often yield findings that fail to generalise. Although heterogeneity, sampling, model dependence and transportability conditions explain particular instances, there is no structural account of why causal precision and generalisability routinely conflict.

\noindent\textbf{Development:} I introduce evidential states as representations of the information available for causal inference at any point in a study. Study operations—restriction (R), conditioning (C) and intervention (I)—are treated as transformations of these evidential states. Each operation removes or reorganises information in different ways. The sequence in which they are applied matters—the operations do not commute. Applying C then R typically yields a different evidential state than applying R then C. Analysing this non-commutativity reveals a structural constraint relating causal precision to evidential breadth.

\noindent\textbf{Application:} In an observational system, conditioning on an observed proxy before restricting the population blocks confounding. Restricting first can eliminate the variation that makes adjustment effective. In an experimental system, randomisation improves internal validity only within the restricted evidential world created by eligibility criteria and post-baseline exclusions. In both cases, the order of operations determines which causal conclusions are supported. These examples illustrate why internally valid findings may fail to transport, why observational adjustments can yield unstable estimates, and why study design choices must be understood as order-dependent transformations of information.

\noindent\textbf{Conclusions:} Causal precision and generalisability cannot be jointly maximised. The operations that secure identification also narrow the evidential world required for transportability. The trade-off is structural, not pragmatic.
\end{abstract}


\maketitle

\begin{tcolorbox}[colback=gray!10,colframe=black,title=Key messages]
	\begin{itemize}
		\item Internal and external validity cannot be jointly maximised.
		\item Routine study operations---restriction, conditioning and intervention---do not commute. The order in which they are applied yields different \textit{evidential states} and can support incompatible causal conclusions.
		\item Non-commutativity imposes an inherent tension between causal precision and generalisability, accounting for familiar difficulties in transportability and external validity.
		\item \textbf{The surer the cause, the smaller the world.}
	\end{itemize}
\end{tcolorbox}

\vspace{1em}

\section{Introduction}

Internal and external validity stand in constant tension in scientific research and this is also true in epidemiology \cite{shadishExperimentalQuasiexperimentalDesigns2002}. The more precisely a study identifies a causal effect, the less the findings apply to a broader populations. Randomised trials with clear effects in homogeneous samples often produce attenuated or inconsistent results in routine clinical settings \cite{rothwellExternalValidityRandomised2005}, and cohort estimates vary across demographic or geographic strata \cite{hernanCausalInferenceWhat2023}. Even well-executed field studies can fail to transport \cite{pearlExternalValidityDoCalculus2014}. These patterns suggest that the very conditions that ensure internal validity frequently undermine external validity.

Several explanations have been given to account for specific failures of generalisation: heterogeneity of treatment effects \cite{kentPersonalizedEvidenceBased2018}, sampling differences \cite{stuartAssessingGeneralizabilityRandomized2015}, model dependence \cite{hoMatchingNonparametricPreprocessing2007}, selection processes \cite{hernanStructuralApproachSelection2004}, and the formal constraints identified by transportability theory \cite{pearlExternalValidityDoCalculus2014}. These explanations clarify when generalisation breaks down but don’t explain why.

The tension is often framed in pragmatic terms, such as limited resources, incomplete measurement, or imperfect sampling \cite{rosenbaumDesignObservationalStudies2020,greenlandBasicMethodsSensitivity1996}. What is missing is a structural explanation; i.e., a way to show how routine study operations themselves generate divergence between internally valid estimates and externally relevant claims, even under ideal conditions.

I argue that the tension arises from the structure of inference itself. Scientific studies transform the evidential world through restriction (R), conditioning (C) and intervention (I). These operations determine which causal claims the resulting data can support. Crucially, the operations do not preserve the same information, nor do they commute. Applying the same components of study design in different sequences (e.g., C then R vs. R then C) yields evidential states that support different causal conclusions.

This order-dependence has direct consequences. Operations that increase causal precision—by narrowing populations or eliminating alternative explanations—reduce the heterogeneity required for external validity. Operations that preserve heterogeneity support generalisability but weaken causal identification. The resulting trade-off is not accidental. It is structural because the evidential transformations that ensures strong causal identification necessarily limit representativeness and variability. The more precisely a study secures a causal estimate, the less representative its evidential world becomes. This constraint applies to observational and experimental studies alike.

The examples that follow illustrate how the tension between identification and generalisation emerges from the way design operations transform the information available for causal inference. By treating restriction, conditioning and intervention as transformations of the evidential world, and by analysing how different sequences preserve or remove variation, we can see how the trade-off occurs.

\section{Development}

Scientific studies do not observe causal structures directly. They record observed patterns which are shaped by design and analytic decisions. The patterns determine which causal explanations remain compatible with the data. To make this explicit, I introduce the notion of an evidential state; that is, a state that represents the information available to an investigator at a given point in a study. It has two components: the empirical distributions observed in the data, and the set of causal models that are consistent with those distributions under the assumptions imposed by the study design.

This construction is deliberately minimalist. It does not require specifying the true data-generating process or the correct causal structure. An evidential state records what the study retains—its remaining variation and dependencies—and what it has ruled out through design decisions such as restricting eligibility or randomising treatment. Each operation reshapes the evidential state. Later operations only act on what earlier ones have left intact. These transformations can remove variation, induce or break dependencies, and constrain the causal models consistent with the data.

\subsection{Study operations}

Epidemiological studies shape the data through a sequence of design and analytic operations. The three fundamental operations are restriction (R), conditioning (C) and intervention (I).

\subsubsection{Restriction (R)}

Restriction encompasses all procedures that limit the population under study. Eligibility criteria, geographic or temporal trimming, exclusion of participants with missing data, and post-baseline selection such as adherence-based per-protocol subsets all count as instances of restriction. They alter the evidential state by removing heterogeneity8,3. The operation, thus, changes the empirical distribution of variables and eliminates some of the pathways through which effects or associations might manifest. This can reduce ambiguity about causal relations—narrower populations often yield cleaner effect estimates—but it also constrains the range of potential effect modifiers, covariate patterns and dependencies that may be present in the broader target population. Restriction alters the observed data and the set of admissible causal models.

\subsubsection{Conditioning (C)}

Conditioning includes familiar analytic operations such as covariate adjustment, stratification, regression control, and matching \cite{greenlandCausalDiagramsEpidemiologic1999,pearlBookWhyNew2020}. It can block confounding paths, restore exchangeability and increase causal precision. It also induces new dependencies and may remove variation that is relevant for generalisation. Whether the operator strengthens or weakens causal identification depends on the structure of the system and on which features of the data remain after earlier design decisions. Conditioning thus transforms the evidential state by altering associations in the data and by modifying which causal explanations remain viable.

\subsubsection{Intervention (I)}

Intervention, in the form of randomisation, assigns treatment independently of pre-treatment covariates within the population defined by prior restrictions. It can eliminate confounding and is often viewed as a gold standard for causal identification. However, intervention does not create representativeness. It operates only on smaller world that restriction has already produced. Moreover, later restrictions---such as excluding non-adherent participants or those lost to follow-up---further restricts, even after randomisation has occurred. Intervention therefore improves internal validity within a constrained domain but cannot by itself restore the heterogeneity required for external validity. 

\subsection{Non-commutativity}

The order of operations (R, C, I) reconfigures the informational conditions under which causal claims can be supported. Restriction followed by conditioning typically produces a different evidential state than performing conditioning followed by restriction.
The reason is straightforward. Each operation removes some features of the data while preserving or inducing others. Restriction removes units and therefore eliminates some of the variation and dependencies present in the original population. Conditioning reorganises the data by fixing or adjusting for certain variables, changing which associations remain visible. Intervention breaks specific dependencies through randomisation, but only within the population that has already been restricted. Because each operation changes the evidential state, later operations act on a different informational substrate than they would have otherwise. Thus, applied in different orders, operations can produce evidential states that support different causal conclusions.
This property is easy to overlook because epidemiological analyses often describe design components—eligibility criteria, covariate adjustment, randomisation—as if they were independent choices that simply accumulate. In virtue of this relationship between the operators, non-commutativity is a structural feature of causal inference.

\subsection{Structural trade-off}

Non-commutativity has direct consequences for causal inference. Operations that narrow the evidential world—by reducing heterogeneity, blocking specific paths or fixing covariates—tend to increase causal precision. They eliminate competing explanations and reduce the range of admissible causal models. By contrast, operations that retain or expand the evidential world—by preserving variation, maintaining heterogeneity or avoiding post-baseline exclusions—support generalisability. They keep open the possibility that findings apply across diverse populations and settings. These two aims, however, require transformations of the evidential state that act in opposing directions.
Narrowing the evidential state to sharpen an effect estimate removes the variability and representativeness required for external validity. Preserving the evidential state to support generalisation retains heterogeneity and dependencies that weaken identification. This follows from the structure of the evidential transformation itself and cannot be laid at the feet of imperfect implementation, incomplete measurement, or unavoidable practical constraints.
The relationship can be expressed schematically by a constraint on what a single evidential state can support:
\[
\Delta_{\mathrm{cause}}(E) \cdot \Delta_{\mathrm{breadth}}(E) \ge k
\]
Here, $\Delta_{\mathrm{cause}}(E)$ reflects how narrowly the data and assumptions restrict the range of plausible causal effects in the evidential state $E$;  $\Delta_{\mathrm{breadth}}(E)$ reflects how far that evidential state has diverged from the full target population in terms of heterogeneity, dependencies and representativeness; and $k$ represents a minimal residual uncertainty arising from the underlying causal structure. The key point is conceptual. The operations that secure strong causal identification ($\Delta_{\mathrm{cause}}(E)$) necessarily transform the evidential state in ways that limit the breadth of claims ($\Delta_{\mathrm{breadth}}(E)$) that can be supported. A formal derivation of the tradeoff for a simple binary system is provided in the Supplementary Material. 

\subsection{An observational example}

Consider the simple causal system shown in Figure 1. The unobserved variable U affects both the treatment T and the outcome Y, creating a confounding path that, if left unblocked, leads to biased estimates of the effect of T on Y. The observed variable X is associated with U and can therefore serve as a proxy. Conditioning on X can block the confounding path and help recover the causal effect. Whether this adjustment succeeds, however, depends on the evidential state produced by earlier operations.

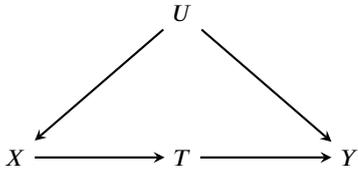
\begin{figure}[ht]
	\centering
	\begin{tikzpicture}[>=stealth, thick]
		\node (U) at (0,0.9) {$U$};        
		\node (X) at (-2.2,-1) {$X$};      
		\node (T) at (0,-1)  {$T$};
		\node (Y) at (2.2,-1)  {$Y$};      
		
		\draw[->] (U) -- (X);
		\draw[->] (U) -- (Y);
		\draw[->] (X) -- (T);
		\draw[->] (T) -- (Y);
	\end{tikzpicture}
	\caption{\textit{Minimal observational causal diagram} illustrating non-commutativity of study operations. An unobserved common cause ($U$) influences both treatment ($T$) and outcome ($Y$). An observed variable ($X$), correlated with $U$, provides a path for adjustment when its variation is preserved. Conditioning on $X$ before restricting the population blocks the confounding path, but restricting first can eliminate variation in $X$, making adjustment impossible. The two sequences of operations lead to different evidential states and incompatible causal conclusions.}
	\label{fig:dag1}
\end{figure}

If we first condition on X in the full population, the remaining variation in X is sufficient to break the association between U and T. The operation retains the blocking structure, and the estimate of the causal effect of T on Y is stable. However, if we restrict the population before conditioning—for example, by excluding individuals with extreme values of X or by selecting a subgroup where X varies only minimally—then conditioning on X no longer blocks the confounding path. The variation in X that previously served to proxy U has been removed. Restriction followed by conditioning yields an evidential state in which the causal effect is not identifiable by adjustment, even though the same operations succeeded when applied in the opposite order.

This simple system shows how non-commutativity arises and how it leads directly to the structural trade-off described above. Conditioning before restriction increases causal precision but reduces generalisability. Restriction before conditioning preserves representativeness but weakens identification.

\subsection{An experimental example}

A similar structure appears in experimental settings, despite the use of randomisation. Figure 2 depicts an idealised trial. The intervention I assigns treatment independently of pre-treatment covariates, thus eliminating confounding within the selected population. This population, however, is already the product of earlier restrictions: eligibility criteria determine who enters the trial. There may also be post-randomisation restrictions, such as exclusions for non-adherence or missing follow-up. The effective sequence of operations is therefore $R \to I \to R'$, where both $R$ and $R'$ modify the empirical distribution and the set of admissible causal models.

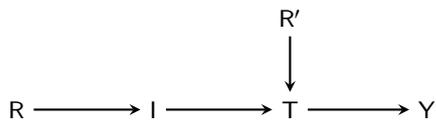
\begin{figure}[ht]
	\centering
\begin{tikzpicture}[>=stealth, thick, node distance=1.4cm]
	\node (R)  at (0,0) {R};
	\node (I)  at (1.8,0) {I};
	\node (T)  at (3.6,0) {T};
	\node (Y)  at (5.4,0) {Y};
	\node (R') at (3.6,1.2) {R$'$};   
	
	\draw[->] (R)  -- (I);
	\draw[->] (I)  -- (T);
	\draw[->] (T)  -- (Y);
	\draw[->] (R') -- (T);            
\end{tikzpicture}	
\caption{\textit{Minimal experimental causal diagram} showing that intervention does not escape evidential non-commutativity. Eligibility criteria ($R$) restrict who is eligible for randomisation ($I$), and post-randomisation restrictions ($R'$)—such as attrition or protocol deviations—determine which treatment assignments ($T$) remain in the analytic sample. Randomisation ensures internal validity within this restricted evidential world, but the restrictions before and after intervention create divergence from the target population. Thus restriction and intervention do not commute, and experimental estimates obey the same causal-generalisability trade-off as observational designs.}
	\label{fig:dag2}
\end{figure}

Randomisation improves internal validity within the restricted evidential world created by these selections. It ensures that, among those included, treatment assignment is independent of measured and unmeasured confounders. However, it does not restore the heterogeneity that restriction has removed. If the eligibility criteria eliminate much of the population variation in key effect modifiers, or if post-baseline exclusions disproportionately remove specific subgroups, the resulting evidential state will diverge substantially from the target population. In this situation, an internally valid causal effect estimate cannot readily generalise.
The non-commutativity of R and I explains this divergence. Restricting the population before randomisation produces a different evidential state than randomising first and then restricting. Because randomisation operates only on the units that survive the initial restriction, reversing the order is not possible in practice, and the information necessary for external validity cannot be recovered by intervention alone. 

\section{Application}

The examples below are not empirical demonstrations but conceptual applications, illustrating how the framework explains common difficulties in generalising causal effects.

\subsection{Failure to generalise}

Randomised trials frequently report effects that attenuate or disappear in routine clinical populations \cite{treweekMakingTrialsMatter2009}. Standard explanations invoke differences in adherence, comorbidities or background risk \cite{zhangAdherenceReportingRandomized2014,hernanPerProtocolAnalysesPragmatic2017}. Within the present framework, these discrepancies reflect the cumulative effects of restriction and intervention on the evidential state. Eligibility criteria reduce heterogeneity in effect modifiers. Randomisation then secures internal validity within that restricted world. Post-baseline exclusions further narrow the domain of inference. Each step increases causal precision and also increases $\Delta_{\mathrm{breadth}}(E)$---the divergence from the broader target population. The operations supports mores precise causal estimate but leave little variation to justify generalisation. 

The constraint is not an artefact of imperfect trial implementation.

\subsection{Transportability}

Transportability analyses formalise the conditions under which causal effects can be moved from one population to another. In practice, these conditions are frequently violated. Trial populations differ from target settings in baseline risk, covariate distributions or effect modifiers, and observational datasets lack precisely the variables required for adjustment \cite{pearlNoteGeneralizabilityStudy2019,pearlExternalValidityDoCalculus2014}. The present framework explains why such violations are pervasive. Operations that achieve internal validity—restriction to adherent participants, control of confounding via conditioning, and stabilisation of treatment delivery—reconfigure the evidential state in ways that eliminate the heterogeneity needed for transport. The information removed by earlier operations cannot be recovered.

\subsection{Study design}

The framework suggests that evaluating study quality requires examining not only the correctness of individual design components but also the sequence in which they are applied. Rather than treating eligibility criteria, covariate adjustment and randomisation as independent choices, investigators should consider how each operation transforms the evidential state and what information is lost as a result. When generalisability is a priority, preserving variation in key covariates and effect modifiers may be more important than maximising internal validity. Conversely, when the goal is precise estimation, accepting a loss of heterogeneity may be unavoidable. The framework does not prescribe a single optimal sequence of operations. It clarifies why the trade-offs recur.

Across these examples, the structural trade-off identified earlier accounts for pervasive patterns in epidemiological research. Internally valid findings fail to generalise because the evidential state has been narrowed through non-commuting operations. Observational findings vary across specifications because conditioning interacts with earlier restrictions in order-dependent ways. Transportability is difficult not only because of practical constraints but because operations that secure causal precision remove the heterogeneity required for transferring effects. These patterns follow from the same mechanism, the order-sensitive transformation of evidential states by restriction, conditioning and intervention.

\section{Discussion}

This paper has argued that the tension between internal and external validity arises from a structural feature of causal inference. Studies transform the evidential world through operations such as restriction, conditioning and intervention. These operations alter the variation, dependencies and admissible causal explanations available in the resulting data. Because the operations change the evidential state in different ways and because their effects depend on the order in which they are applied, they do not commute. Non-commutativity imposes a structural constraint. Evidential states that support precise causal estimates are not the same as those that support generalisable claims. 

The framework provides a unified account of several longstanding methodological puzzles. In addition to commonly cited issues of adherence, comorbidity or care delivery, internally valid effects frequently fail to generalise because the evidential world in which they were identified has been narrowed by design decisions that eliminate the heterogeneity required for transportability. While incomplete data or practical constraints affect transportability, it also fails because the very operations that strengthen causal identification remove the variation needed for results to hold across populations. These phenomena have typically been treated as distinct problems. The present framework, however, shows that they share a common structural source.

The main value of the account is conceptual clarity. By making the evidential state explicit and analysing how study operations transform that state, it becomes possible to see why one cannot satisfy both high internal validity and wide external validity. The framework does not prescribe an optimal study design, but it clarifies the trade-offs inherent in choosing among them. In settings where generalisability is essential, investigators may prioritise heterogeneity, minimising stringent eligibility criteria or avoiding analytic decisions that remove variation in key effect modifiers. Conversely, when the aim is the precise estimation of a causal, a narrower evidential world may be appropriate, with the understanding that generalisability is sacrificed. Recognising the structural basis of the trade-off can help align design decisions with the intended scope of inference.

Several limitations should be acknowledged. The account is intentionally minimal. It focuses on three core operations and on small causal systems used for illustration. Real-world studies involve more complex designs, time-varying exposures, feedback processes and non-standard sampling frames. The framework misses that complexity. Instead, it offers a conceptual tool for reasoning about how design choices interact and why certain trade-offs appear even in simplified settings. Future work could extend the framework to longitudinal designs, incorporate formal metrics of representativeness or explore how alternative study operations (e.g., imputation, weighting or model-based standardisation) transform evidential states. Empirical applications may also identify settings where the structural constraint binds more tightly, or where particular design sequences preserve generalisability more effectively.

Despite the limitations, the core claim holds. Causal specificity and generalisability cannot be maximised together. Treating design choices as order-dependent transformations of evidential states explains why. A succinct maxim of this structural constraint is, ``the surer the cause, the smaller the world''.

\section{Funding}

Not applicable

\section{Acknowledgement}

Three artificial intelligence models (ChatGPT 5.1 (OpenAI), Claude Sonnet 4.1 (Anthropic), and Kimi 2 (Moonshot AI)) were used extensively as editorial aids in the drafting and redrafting of this paper. The tools were used to check the logic, flow and coherence of the argument, tighten expressions, and reduce word count. Interchanging their use provided different “editorial voices”, and independent checks on the logic and flow of arguments. They were also used for checking the supplementary material and code correcting the \LaTeX{} layout.

\appendix
\section*{Supplementary Information}

\subsection*{S1. Evidential states and inferential operations}

Let an \emph{evidential state} be
\[
E = (P,\mathcal{M}),
\]
where \(P\) is the joint distribution over observed variables and
\(\mathcal{M}\) is the set of causal models compatible with \(P\).

Three inferential operations act on \(E\):

\begin{enumerate}
	\item \textbf{Conditioning} on an observed variable \(X\):
	\[
	C(E) = \big(P(\cdot\mid X),\,\mathcal{M}_C\big), \qquad 
	\mathcal{M}_C \subseteq \mathcal{M}.
	\]
	
	\item \textbf{Restriction} to a subpopulation defined by \(S\):
	\[
	R(E) = \big(P(\cdot\mid S),\,\mathcal{M}_R\big), \qquad 
	\mathcal{M}_R \subseteq \mathcal{M}.
	\]
	
	\item \textbf{Intervention} on a treatment variable \(T\), applied only
	to the units that survive earlier restrictions:
	\[
	I(E) = \big(P_{do(T=t)},\,\mathcal{M}_I\big), \qquad
	\mathcal{M}_I \subseteq \mathcal{M}.
	\]
\end{enumerate}

Each operation contracts both the empirical distribution and the admissible model set.

\subsection*{S2. Minimal binary system}

Consider the following binary causal system:

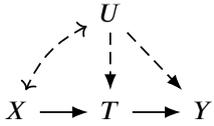
\begin{tikzcd}[column sep=1.8em, row sep=2.2em, every arrow/.append style={-Latex}]
	& U \ar[dl, dashed, <->, bend right=20] \ar[d, dashed] \ar[dr, dashed] & \\
	X \ar[r] & T \ar[r] & Y
\end{tikzcd}

\vspace{0.75em}

It consists of a treatment $T$, outcome $Y$, observed proxy $X$, and unobserved confounder $U$. $X$ is correlated with $U$ and thus serves as a proxy that can block the backdoor path.

Let all variables be in \(\{0,1\}\), and write
\[
T = f(U,X,\varepsilon_T), \qquad 
Y = g(T,U,\varepsilon_Y).
\]

The target causal effect is
\[
\tau = \mathbb{E}[Y \mid do(T=1)] - \mathbb{E}[Y \mid do(T=0)].
\]

This system is minimal in the sense that conditioning can identify \(\tau\), but restriction may destroy the variation that makes conditioning effective.

\subsection*{S3. Non-commutativity of inferential operations}

\subsubsection*{S3.1 Conditioning then restriction}

Conditioning on \(X\) before restricting preserves sufficient variation in X for adjustment. Since \(X\) blocks confounding,
\[
\tau \text{ is identifiable in } CR(E).
\]
The resulting model set is
\[
\mathcal{M}_{CR}
= \{\, M\in \mathcal{M} : X\text{ blocks the backdoor path} \,\}.
\]

\subsubsection*{S3.2 Restriction then conditioning}

Suppose the restriction selects a subset \(S\) for which
\[
\mathrm{Var}(X\mid S) = 0.
\]
Then conditioning on \(X\) yields no adjustment:
\[
\tau \text{ is not identifiable in } RC(E).
\]
The model set becomes
\[
\mathcal{M}_{RC}
= \{\,M\in \mathcal{M} : \tau \text{ underdetermined from } P(\cdot\mid S)\,\}.
\]

\subsubsection*{S3.3 Non-commutativity}

Because both \(P\) and \(\mathcal{M}\) differ across sequences,
\[
CR(E) \neq RC(E).
\]

\subsubsection*{S3.4 Quantities defining the causal–breadth constraint}

\paragraph{Causal precision.}
Treat \(\tau\) as a random variable over the admissible model set \(\mathcal{M}_E\).
Define
\[
\Delta_{\mathrm{cause}}(E)
=
H(\tau \mid E_{\mathrm{prior}})
-
H(\tau \mid E),
\]
the entropy reduction attributable to evidential state \(E\).

\paragraph{Evidential breadth.}
Define
\[
\Delta_{\mathrm{breadth}}(E)
=
D_{\mathrm{KL}}\!\big(P_E \,\|\, P_{\mathrm{full}}\big),
\]
the divergence between the empirical distribution under \(E\) and that of the full population.%
\footnote{The assumption is that \(P_{\mathrm{full}}\) dominates \(P_E\), ensuring the KL divergence is finite. In practice, this means the restricted state corresponds to a subset of the full population.}

CR increases \(\Delta_{\mathrm{cause}}\) on average but enlarges the divergence \(\Delta_{\mathrm{breadth}}\).  
RC preserves breadth but weakens precision.

\subsubsection*{S3.5 Intervention and the persistence of the constraint}

Intervention is never applied to the full evidential state but only to a restricted version:
\[
E \;\longrightarrow\; R_1 \;\xrightarrow{\ I\ }\; R_2(E),
\]
where \(R_1\) is eligibility restriction and \(R_2\) represents post-intervention selection (attrition, non-adherence, protocol deviations).

\paragraph{(a) Randomisation increases causal precision.}
Random assignment collapses admissible models within the restricted world, increasing \(\Delta_{\mathrm{cause}}\).

\paragraph{(b) Restrictions contract evidential breadth.}
Both \(R_1\) and \(R_2\) increase divergence from the full distribution:
\[
D_{\mathrm{KL}}(P_{R_1IR_2} \,\|\, P_{\mathrm{full}}) > 0.
\]

\paragraph{(c) Non-commutativity.}
Because effect-modifier variation depends on order,
\[
RIR(E) \neq IRR(E).
\]

\paragraph{(d) Causal–breadth constraint.}
The product
\[
\Delta_{\mathrm{cause}}\!\big(R_1IR_2(E)\big)\;
\Delta_{\mathrm{breadth}}\!\big(R_1IR_2(E)\big)
\]
is bounded below by a constant \(k\).

\paragraph{(e) Consequence.}
An RCT identifies \(\tau\) only within the restricted evidential world in which randomisation occurred.  
Generalisability requires heterogeneity that restrictions may eliminate.

\subsection*{S4. Derivation of the lower bound \(k\)}

We assume throughout that the causal effect \(\tau\) is only partially identified from the observational distribution, i.e. the entropy \(H(\tau \mid M)\) is strictly positive for every model \(M \in \mathcal{M}_{\mathrm{full}}\). Let
\[
\mathcal{T}(E)
=
\{\tau(M): M\in\mathcal{M}_E\}
\]
be the set of admissible causal effects under state \(E\).

Define the minimal residual uncertainty
\[
k
=
\inf_{M\in\mathcal{M}_{\mathrm{full}}}
H(\tau \mid M).
\]
If \(\tau\) is point-identified in the full population, then \(H(\tau\mid M)=0\) for some model and thus \(k=0\).  
The constraint is active only when \(\tau\) is partially identified, i.e.\ when \(H(\tau\mid M)>0\) for all \(M\in\mathcal{M}_{\mathrm{full}}\).

For any contraction operator \(O\in\{C,R,I\}\), the data-processing inequality implies:
\begin{align*}
	H(\tau \mid OE) &\le H(\tau \mid E), \\
	D_{\mathrm{KL}}\!\big(P_{OE}\,\|\,P_{\mathrm{full}}\big) 
	&\ge D_{\mathrm{KL}}\!\big(P_E\,\|\,P_{\mathrm{full}}\big).
\end{align*}
Hence
\[
\Delta_{\mathrm{cause}}(OE)\;
\Delta_{\mathrm{breadth}}(OE)
\;\ge\;
k,
\]
with equality iff $OE$ is sufficient for point identification of $\tau$ and $P_{OE}=P_{\mathrm{full}}$.

\subsection*{S5. Interpretation}

\begin{itemize}
	\item \(k\) increases with heterogeneity and latent confounding.
	\item \(k\) approaches zero when the causal effect is already identifiable in the full system.
	\item The constraint tightens when identification depends on variation that restrictions erase.
\end{itemize}

\subsection*{S6. Summary}

The causal–breadth constraint
\[
\Delta_{\mathrm{cause}}(E)\ \cdot \Delta_{\mathrm{breadth}}(E) \ge k
\]
follows from the non-commutativity of conditioning, restriction, and intervention.  
Causal precision increases only through contraction of the evidential world, and this contraction inevitably limits generalisability.

\bibliographystyle{cas-model2-names}
\bibliography{CUP_251127}

\end{document}